\documentclass[fleqn,10pt]{wlscirep}
\usepackage{amsmath}
\usepackage{amsfonts}
\usepackage{amssymb}
\usepackage{graphicx}
\usepackage{gensymb}
\usepackage[colorlinks=true]{hyperref}

\title{Multistep transition of diamond to warm dense matter state revealed by femtosecond X-ray diffraction}
\author[1,2,*]{Nikita Medvedev} 
\author[3,4,**,+]{Beata Ziaja} 

\affil[1]{Institute of Physics, Czech Academy of Sciences, Na Slovance 2, Prague 8, 18221, Czech Republic} 
\affil[2]{Institute of Plasma Physics, Czech Academy of Sciences, Za Slovankou 3, Prague 8, 18200, Czech Republic} 
\affil[3]{Center for Free-Electron Laser Science, DESY, Hamburg, 22607, Germany} 
\affil[4]{Institute of Nuclear Physics, Polish Academy of Sciences, Krak\'{o}w, 31-342, Poland} 

\affil[*]{Corresponding author: nikita.medvedev@fzu.cz} 
\affil[**]{Corresponding author: ziaja@mail.desy.de} 
\affil[+]{The authors contributed equally to this work} 

\keywords{Diamond, Warm Dense Matter, X-ray Free Electron Laser, XTANT}

\begin{abstract} 
Diamond bulk irradiated with a free-electron laser pulse of 6100 eV photon energy, 5 fs duration, at the $\sim 19-25$ eV/atom absorbed doses, is studied theoretically on its way to warm dense matter state. Simulations with our hybrid code XTANT show disordering on sub-100 fs timescale, with the diffraction peak (220) vanishing faster than the peak (111). The warm dense matter formation proceeds as a nonthermal damage of diamond with the band gap collapse triggering atomic disordering. Short-living graphite-like state is identified during a few femtoseconds between the disappearance of (220) peak and the disappearance of (111) peak. The results obtained are compared with the data from the recent experiment at SACLA, showing qualitative agreement. Challenges remaining for the accurate modeling of the transition of solids to warm dense matter state and proposals for supplementary measurements are discussed in detail. 
\end{abstract}

\begin{document} 

\flushbottom 
\maketitle 

\thispagestyle{empty} 

%-------------------------------------------------------- 
\section*{Introduction} 

Over the years, extensive theoretical studies of many-body systems lead to a development of two complementary approaches towards their description: 
(i) In case when the average kinetic energy of constituent particles is much larger than the average potential energy between them and than the Fermi energy of the system, the system is in the classical, ideal plasma state \cite{Murillo1998}. This state can be well described within semi-classical approximations \cite{Zwanzig2001,Birdsall2005}. 
(ii) In the opposite case, when the Fermi energy and the average potential energy are much larger than the average kinetic energy of the particles, the system is in the condensed matter state. For its description, the quantum methods including the description of interparticle interactions should be applied. In particular, the density-functional theory (DFT) methods have been extensively developed for the last few decades \cite{Parr1980,Marques2012}. 

At moderate temperatures and densities, when the two regimes (i) and (ii) meet, a new state, the so-called warm dense matter (WDM) emerges \cite{Graziani2014, Redmer2008, Valenza2016}. Neither of the above mentioned approaches can be rigorously applied in this specific regime, as the potential energy of interaction among the particles is of the same order as their kinetic energies. No rigorous theory or model has been proposed so far to describe properties and behavior of the exotic state of matter, in particular, the non-equilibrium transition from condensed matter into the plasma state. This is one of the most challenging problems bridging the solid-state and the plasma communities \cite{Graziani2014}.

Yet, the WDM state is common in the Universe: it exists in the inner core of large planets (such as Jupiter and Saturn), in white dwarf stars, and, supposedly, on the surface of neutron stars \cite{Graziani2014, Redmer2008}. In the laboratories, it is produced as a transient state, following a deposition of high energy dose into a solid target \cite{Graziani2014, Fletcher2015}. Such energy deposition is often performed through the irradiation of the target sample with high-intensity ultrashort laser pulses, e.g., strongly focused X-ray pulses from the free-electron lasers (FEL) such as FLASH \cite{Ackermann2007}, LCLS \cite{Bostedt2016}, SACLA \cite{Pile2011}, SwissFEL \cite{Ganter2011}, and European XFEL \cite{Altarelli2011}. The modern free-electron lasers provide femtosecond intense pulses of X-ray photons, sufficiently bright to create a WDM state in a single shot \cite{Ackermann2007, Bostedt2016, Pile2011}. These new experimental opportunities create a strong demand for a theoretical support and numerical modeling to describe experimental results, to suggest and guide new experiments, and ultimately to understand the fundamental physics of the warm dense matter.

In an ultrafast strongly driven matter, intrinsic processes occur at femtosecond timescales \cite{Lorazo2006, Zastrau2012, Medvedev2015}. Excited electrons redistribute their energy among themselves and, later, also transfer it to the ions. The latter process may lead to material modifications, and to formation of transient nonequilibrium plasma \cite{Ziaja2008}. 

In the recent experiment at the free-electron laser facility SACLA, diamond was studied with an X-ray X-ray pump-probe scheme \cite{Inoue2016}. The experiment utilized the modern capability of the FELs: two color mode, enabling almost homogeneous deposition of an extremely high fluence into the sample with the pump pulse, and a measurement of the X-ray diffraction patterns with the probe pulse. The probe pulse arrived after a certain femtosecond delay from the pump pulse. The photon energy of the pump pulse was 6100 eV (2.03 \AA), whereas the probe pulse had photon energy of 5900 eV (2.10 \AA). Both pulse durations were 5 fs \cite{Inoue2016}. 

The pump fluences achieved in the experiment were $2.3 \times 10^4$ J/cm$^2$, $2.7 \times 10^4$ J/cm$^2$, and $3.1 \times 10^4$ J/cm$^2$. For 6.1 keV X-rays, such fluences result in average absorbed doses of 18.5 eV/atom, 21.7 eV/atom, and 24.9 eV/atom respectively, deposited in diamond within the photon attenuation length of 279 $\mu$m \cite{Henke1993}, much larger than the sample thickness. After such energy deposition, the material turns into warm dense matter state \cite{Zastrau2012,Dharma-wardana2016a}. 

Monitoring X-ray diffraction patterns of irradiated diamond, Inoue {\em et al.} observed that the integrated probe diffraction intensity of the (220) reflection decreased faster than that of the reflection (111). They both significantly decreased within 80 fs after the maximum of the pump pulse, which was the longest delay available between the pump and the probe pulses in that experiment \cite{Inoue2016}. 

The interpretation of the results turned out to be challenging. The decreasing intensity of the diffraction peaks could be analyzed in Ref.~[\citeonline{Inoue2016}] only in the framework of the Debye-Waller factors, assuming thermal atomic oscillations. A possibility of nonthermal material destabilization due to electronic excitation was also discussed by the authors. Such nonthermal phase transition results from a change in the potential energy surface of atoms due to the electronic excitations. A well known example of such transition is nonthermal melting, see, e.g., \cite{Siders1999, Sundaram2002}. In contrast, thermal phase transition is due to the kinetic energy exchange between hot electrons and atoms (e.g., electron-phonon coupling) through non-adiabatic coupling between the two subsystems. 

However, the contribution of those mechanisms was addressed in Ref.~[\citeonline{Inoue2016}] only on the level of a general discussion, not supported by any quantitative theoretical analysis.  It remained then unclear which physical mechanism did lead to the experimental observation, and why one of the peaks decreased faster than the other. These questions are of the main interest for the current study which uses a modeling tool that can address transient processes occurring on sub-ps timescale.

We apply the recently developed hybrid code XTANT (X-ray-induced Thermal And Nonthermal Transitions~\cite{Medvedev2013e,Medvedev2017}) to model X-ray irradiated diamond at the experimental conditions used in Ref.~[\citeonline{Inoue2016}]. The model consists of a few modules dedicated to simulate various processes induced by the incoming X-ray FEL radiation: 

(a) The core of the model is the transferable tight binding (TB) Hamiltonian, which treats electronic band structure and atomic potential energy surface. It evolves in time, depending on the positions of all the atoms in the simulation box. 

(b) Atomic positions are propagated in time using a classical molecular dynamics (MD) scheme. It solves Newton equations for nuclei, with the interaction potential evaluated within the TB module. 

(c) Electron occupation numbers within the TB-based band structure are assumed to follow Fermi-Dirac distribution with a transient temperature and chemical potential evolving in time. The electron temperature is changing due to interaction of band electrons with X-rays, high-energy electrons excited by X-ray photons, impact ionizations, or Auger-decays; or due to their nonadiabatic interaction with nuclei (through electron-phonon, or more generally, electron-ion scattering). 

(d) Non-equilibrium part of the electron distribution at high-energies is treated with a classical event-by-event Monte Carlo simulation. It stochastically models X-ray induced photoelectron emission from K-shell or from the valence band, Auger decays, and the scattering of high-energy electrons.

(e) Electron-ion energy exchange, mentioned above, is calculated with a nonadiabatic approach, in which matrix elements are calculated as an overlap of TB wave-functions plugged into the Boltzmann collision integral.

More details are given in the Methods section. This combined approach allows to treat predominant processes within an X-ray FEL irradiated samples, including their non-equilibrium evolution stage, possible nonthermal effects and phase transitions \cite{Medvedev2013f}. 
With this approach, we will now address the formation of warm dense matter from an X-ray irradiated diamond under the conditions of the experiment \cite{Inoue2016}. 

\section*{Results}

\begin{figure}[th!] 
\centering 
\includegraphics[width=0.94\linewidth,trim={0 0 0 0},clip]{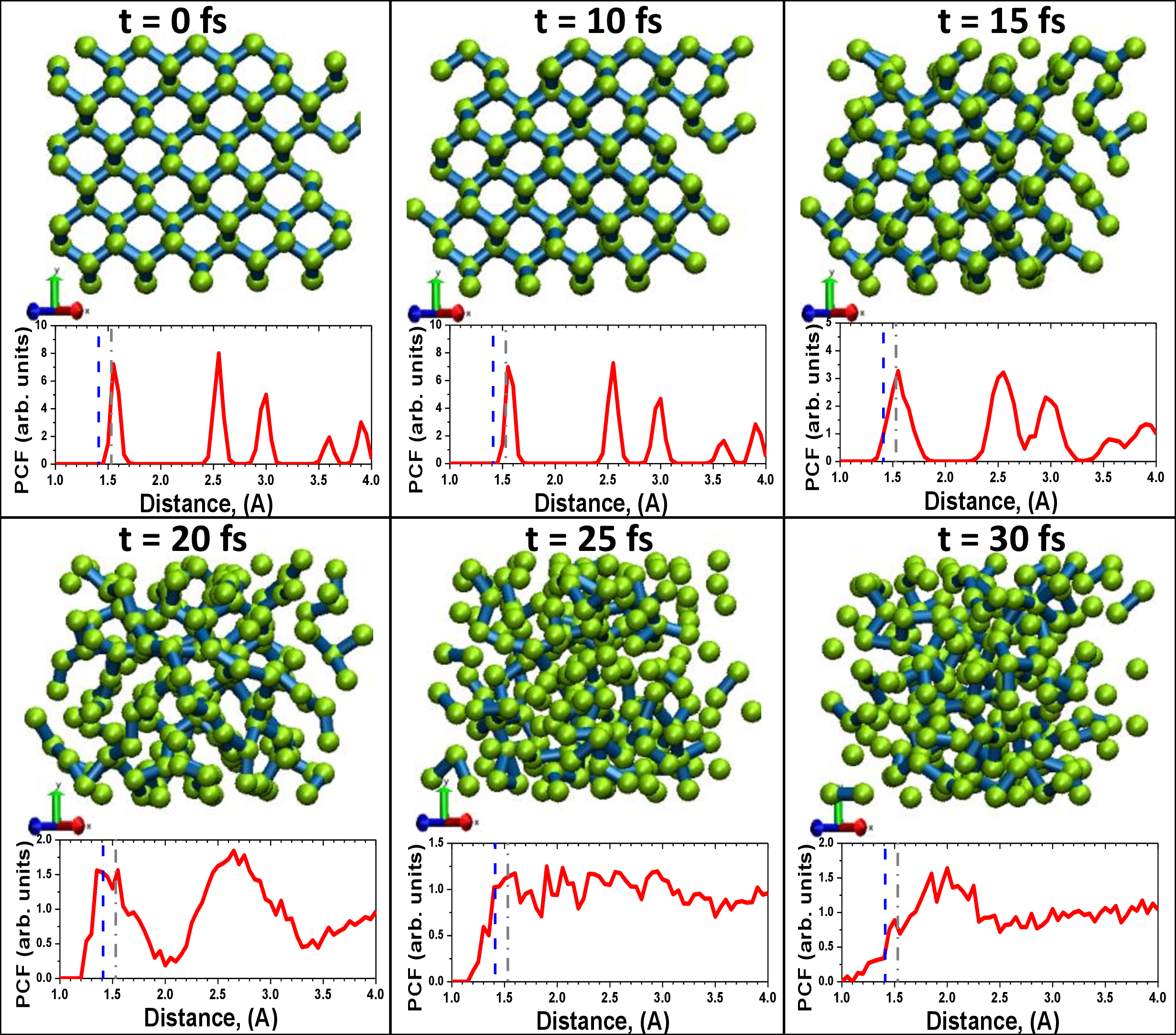} 
\caption{Atomic snapshots of diamond irradiated with X-ray pulse of 6.1 keV photon energy, 5 fs FWHM duration. The pulse deposits an average absorbed dose 24.9 eV per atom. Insets show a pair correlation function (PCF) at the corresponding time instants. Gray dash-dotted vertical line in the insets mark the nearest neighbor distance in diamond, whereas blue dashed lines indicate the nearest neighbor distance in graphite.} 
\label{fig:Snapshots} 
\end{figure}

The simulation of diamond irradiated with an X-ray FEL pulse (Gaussian pulse of 5 fs FWHM duration; 6.1 keV photon energy; average absorbed dose 24.9 eV/atom) performed with XTANT code demonstrates that the atomic structure quickly disorders, on a timescale of a few tens of femtoseconds, see atomic snapshots in Fig.~\ref{fig:Snapshots}. This damage proceeds as a sequence of steps, similar to the graphitization of diamond occurring at lower fluences (reported earlier in, e.g., Ref.~[\citeonline{Gaudin2013}] and confirmed in Ref.~[\citeonline{Tavella2017}]), however, on much shorter time-scales. During the simulation, firstly, valence electrons are excited into the conduction band during and for some time after the pulse. The number of the excited electrons is high, overcoming the damage threshold value ($\sim 1.5$ \%)\cite{Medvedev2013e} already early during the exposure to the pump pulse \cite{Gaudin2013}. This leads to the band gap collapse at the time $\sim 15$ fs. This evolution stage is documented in more detail in the Methods section.

\begin{figure}[h] 
\centering
\includegraphics[width=0.94\linewidth,trim={20 0 30 0},clip]{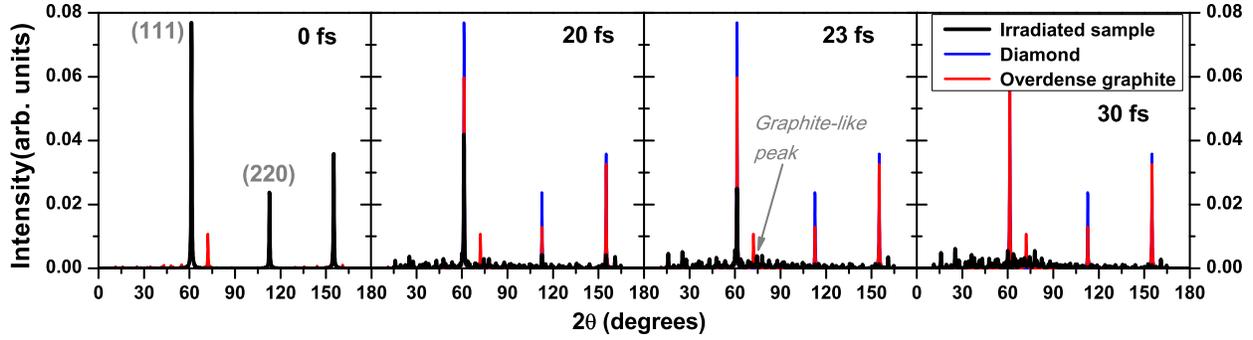} 
\caption{Powder diffraction patterns in diamond irradiated with an X-ray pulse of 6.1 keV photon energy, 5 fs FWHM duration, at the average absorbed dose of 24.9 eV/atom at different time instants after the pump pulse maximum.} 
\label{fig:Diffraction_snapshots} 
\end{figure}

Diamond transiently undergoes the graphitization stage, lasting only for a few femtoseconds. It is indicated  by a transient presence of the graphite nearest-neighbor peak in the pair correlation function in Fig.~\ref{fig:Snapshots}, and a small peak in the diffraction patterns (Fig.~\ref{fig:Diffraction_snapshots} at 20 fs and 23 fs; the intensities of diffraction peaks from our simulated atomic snapshots were extracted using an open software reciprOgraph\cite{reciprOgraph}). Further material disordering continues from this graphite-like phase. A quick atomic rearrangement follows which leads to the sample disordering at times $>20$ fs, see Fig.~\ref{fig:Snapshots}. It is due to the fact that the radiation dose delivered to the material is sufficiently high not only to graphitize diamond, but to also damage the forming graphite on femtosecond timescale. In particular, the previously reported ablation dose in graphite ($\sim 3.3$ eV/atom~\cite{Jeschke2001}) is overreached already during the beginning of the FEL pulse, triggering a rapid damage of the graphite as soon as it is formed.

\begin{figure}[h] 
\centering 
\includegraphics[width=0.5\linewidth,trim={20 10 20 35},clip]{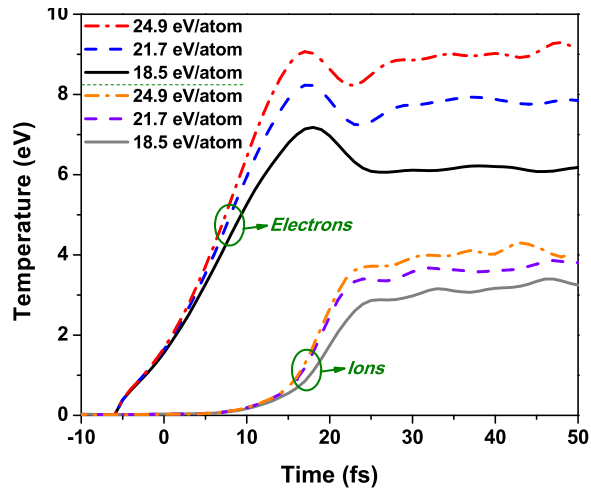} 
\caption{Temperatures of electrons and ions in diamond irradiated with an X-ray pulse of 6.1 keV photon energy, 5 fs FWHM duration, at the average absorbed doses of 18.5 eV/atom, 21.7 eV/atom, 24.9 eV/atom. They correspond to the experimental fluences of $2.3 \times 10^4$ J/cm$^2$, $2.7 \times 10^4$ J/cm$^2$, and $3.1 \times 10^4$ J/cm$^2$ respectively.} 
\label{fig:Temperatures} 
\end{figure}

This process is clearly of nonthermal nature at its early stage, i.e., until $15-20$ fs. The temperatures of electrons and ions are shown in Fig.~\ref{fig:Temperatures}. It is important to emphasize that the ion temperature increases here due to nonthermal changes of the potential energy surface, and not due to electron-ion (electron-phonon) energy exchange. 

\begin{figure}[h] 
\centering 
\includegraphics[width=0.5\linewidth,trim={20 
10 20 30},clip]{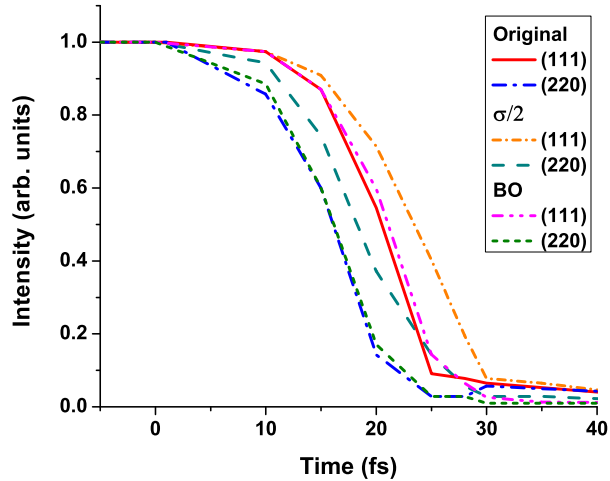} 
\caption{Integrated diffraction peak intensities (111) and (220) in diamond irradiated with an X-ray pulse of 6.1 keV photon energy, 5 fs FWHM duration, at the absorbed dose of 24.9 eV/atom ($3.1 \times 10^4$ J/cm$^2$). The three cases are presented: full calculations, Born-Oppenheimer calculations (BO, electron-ion coupling excluded), and calculations with artificially reduced electron inelastic scattering cross sections (marked as $\sigma/2$).}
\label{fig:Test_peaks} 
\end{figure}

To justify this statement, we performed a simulation within Born-Oppenheimer approximation that naturally excludes nonadiabatic electron-ion coupling. Fig.~\ref{fig:Test_peaks} compares the full calculations with the BO approximation. The BO-results are nearly identical to the case with electron-ion coupling included. This confirms a negligible contribution of electron-ion coupling at such extremely short timescales, and proves that the ion temperature increase and the material damage are of nonthermal origin.

Electronic ensemble reaches temperatures of 7 to 9 eV at the time of $\sim 20$ fs after the maximum of the pump pulse, due to photoabsorption and further secondary high-energy electron scattering. Such an extreme level of electronic excitation severely modifies potential energy surface that keeps ion lattice stable \cite{Medvedev2013e}. Ions remain cold until $\sim 15$ fs. Their temperature starts to raise quickly when the modified interatomic potential pushes them out from their former equilibrium positions, at the same time increasing their temperature up to $\sim 3$ eV. The timescale of the ion temperature increase is in agreement with the estimates of ion displacements obtained with the Debye-Waller fit in Ref.~[\citeonline{Inoue2016}] (Fig.~4 therein). The predicted displacements start to increase fast at the times above 20 fs, in agreement with our predictions on ion temperature from Fig.~\ref{fig:Temperatures}. This way, a warm dense matter state is formed on a ten-femtosecond timescale.

To support the proposed mechanism of diamond's structural disordering, we will now compare our simulation results to the experimental data. In Ref.~[\citeonline{Inoue2016}], the authors presented detailed data on the decrease of (111) and (220) diffraction peak intensities in diamond after its irradiation with an X-ray FEL pulse.

\begin{figure}[h] 
\centering 
\includegraphics[width=\linewidth,trim={20 0 40 0},clip]{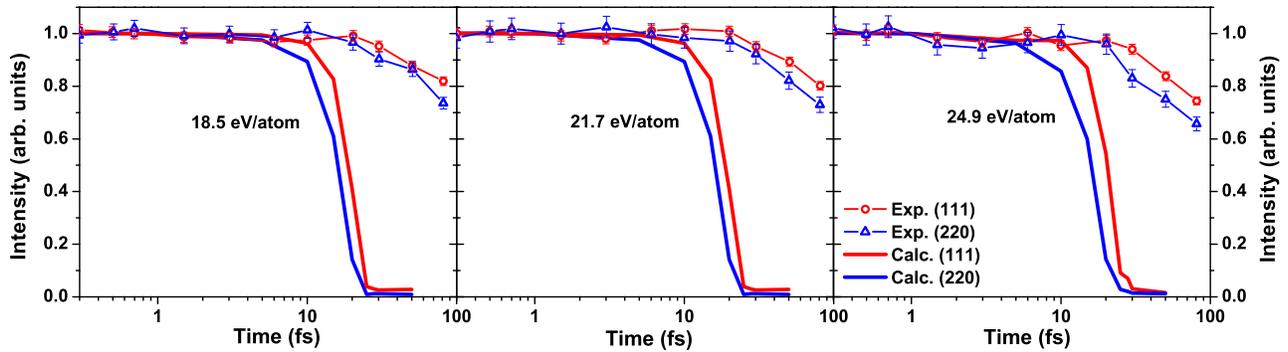} 
\caption{Integrated diffraction peak intensities (111) and (220) in diamond irradiated with an X-ray pulse of 6.1 keV photon energy, 5 fs FWHM duration, at the absorbed doses of 18.5 eV/atom (left panel), 21.7 eV/atom (middle panel), 24.9 eV/atom (right panel). They correspond to the experimental fluences of $2.3 \times 10^4$ J/cm$^2$, $2.7 \times 10^4$ J/cm$^2$, and $3.1 \times 10^4$ J/cm$^2$ respectively. The experimental data are depicted with lines with markers, whereas the calculated results are solid lines.} 
\label{fig:Peaks} 
\end{figure}

The calculated results in Fig.~\ref{fig:Peaks} show qualitative agreement with the experiment. The intensity of peak (220) decreases faster than the peak (111) both in our simulation and in the experiment. However, quantitatively, the simulated diffraction peaks vanish significantly faster (by the time of $\sim 25$ fs) than the experimental ones ($> 80$ fs). The reasons for this discrepancy will be discussed later.

\begin{figure}[h] 
\centering 
\includegraphics[width=0.7\linewidth,trim={15 
220 170 15},clip]{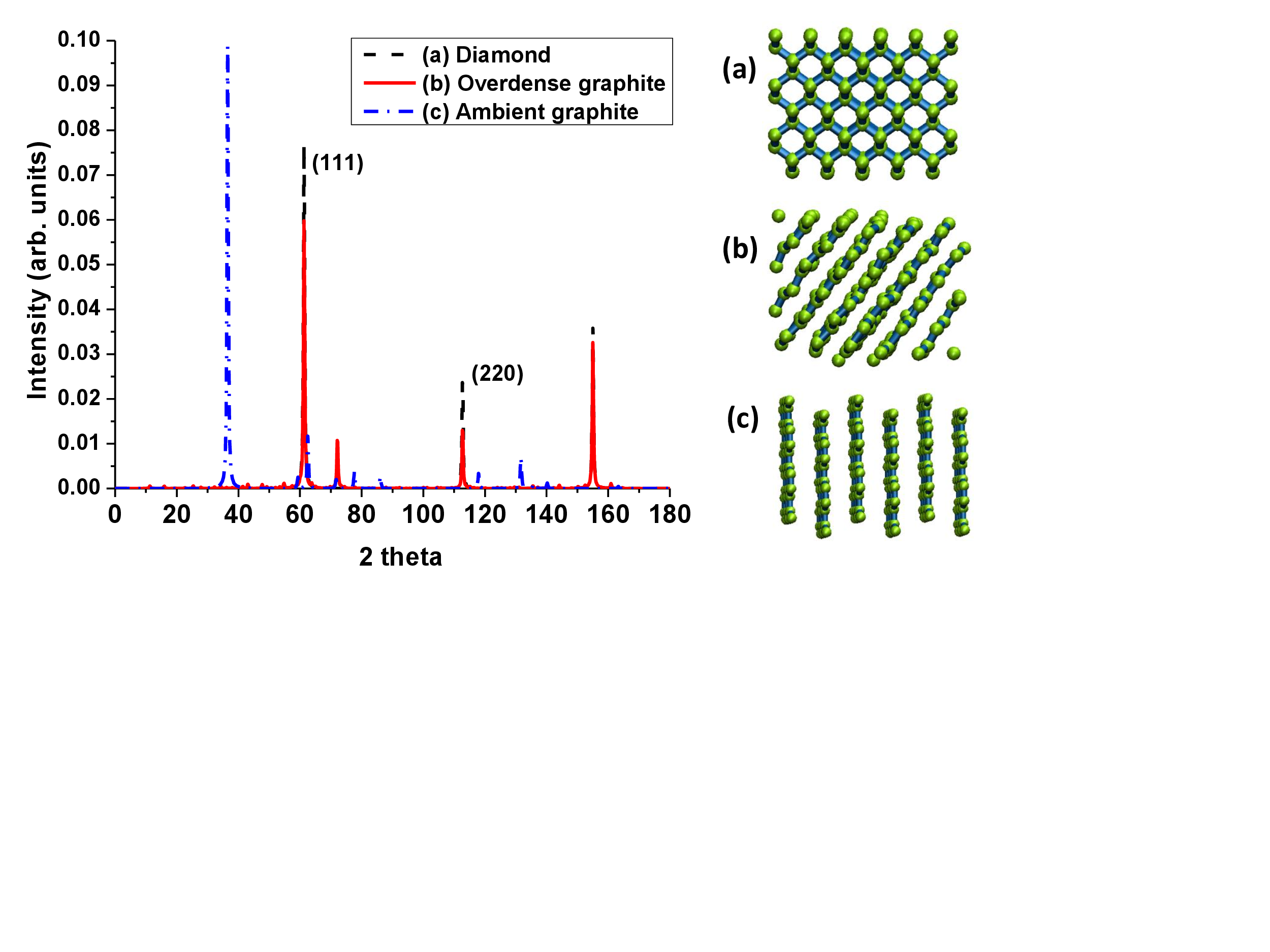} 
\caption{Diffraction peaks of (a) diamond, (b) overdense graphite (with density equal to that of diamond), and (c) ambient graphite at the wavelength of 2.1 \AA; left panel. Corresponding atomic structures are shown in the insets on the right.} 
\label{fig:Equilibrium_diffraction_snapshots} 
\end{figure}

To interpret those findings, we first discuss the features of the diffraction peaks of equilibrium diamond, overdense and equilibrium graphite. Fig.~\ref{fig:Equilibrium_diffraction_snapshots} shows the diffraction peaks at the probe wavelength of 2.1 \AA\ (5900 eV). The corresponding atomic structures are shown in the insets on the right. One can note that the intensity of the diamond reflection (111)  is only slightly reduced in the overdense graphite (with the density equal to that of diamond, as transiently formed during the graphitization process before material expansion), whereas the intensity of the (220) reflection is reduced by half. The reason is that the changes of the (220)-peak indicate a change of the nearest neighbor distance in diamond which is a primary indication of a progressing structural transition. Thus, the strong reduction of the (220)-peak, with the (111)-peak still present, indicates a progressing graphitization process.

Indeed, in the experiment the intensity of the reflection (220) decreases faster than of the reflection (111). This indicates that the damage proceeds via a transient graphitization phase, similarly to that described in Ref.~[\citeonline{Medvedev2013f}]. %During this phase, overdense graphite diffraction peaks are emerging for a few femtoseconds. 
After that, the graphite-like structure disorders, which leads to a disappearance of the (111) peak. The simulation results indicate that the femtosecond time delay between the disappearance of the peak (220) and the peak (111) corresponds to the duration of the transient graphite-like state. This prediction can be tested in the  future provided that the temporal resolution of pump-probe experiments improves enough to measure such short-living states.

Additionally, Ref.~[\citeonline{Inoue2016}] presents angular positions of the diffraction peaks. Only slight shifts ($<0.15\degree$) of the maxima positions for both (111) and (220) peaks towards smaller angles were observed at the timescale of 80 fs. Although the absorbed dose exceeded: (i) the threshold for the graphitization of diamond, $\sim 0.7$ eV/atom, corresponding to $1.5$ \% of excited valence electrons, (ii) the threshold for damage of graphite, $\sim 2$ eV/atom, corresponding to $\sim 9$ \% of excited electrons \cite{Jeschke2001}, and (iii) the ablation threshold of carbon, $\sim 3.3$ eV/atom, corresponding to $\sim 12$ \% of excited valence electrons \cite{Jeschke2001}, estimated with XTANT, the lack of any significant shifts of maxima positions of both Bragg peaks indicates that the material expansion due to ablation was insignificant at the timescale of 80 fs. It justifies the usage of MD simulation scheme at a constant volume (V=const, NVE ensemble) which we applied here.

%-------------------------------------------------------- 
\section*{Discussion} 

Our results achieve only qualitative agreement with experiment. In particular, the simulated timescales of WDM formation are much shorter than those observed in experiment~\cite{Inoue2016}. This is due to a few limitations of the XTANT model, which manifest here due to the extreme irradiation conditions of the simulated experiment. Below we discuss in detail the reasons for the discrepancy observed.

Firstly, it is to be expected that the created high-energy photo-electrons are, in reality, cascading for longer times than estimated with the model. The XTANT code currently uses cross section for electron impact ionization, calculated for neutral diamond from the complex dielectric function \cite{Medvedev2013e,Medvedev2015}. The calculations with different impact ionization cross sections for neutral sample produce very similar cascading times \cite{Medvedev2015cc}. However, here, due to a strong ionization of the sample by a high radiation dose deposited, the sample neutrality quickly breaks down. There are no rigorously derived impact ionization cross sections in highly excited solids yet available. Nevertheless, it can be expected that the cross sections calculated within highly excited solids should be smaller than those obtained for neutral samples, as the ionized nuclei attract the valence electrons stronger, and their binding energies would increase as already shown for plasmas, e.g., in Ref.~[\citeonline{Son2014}]. Including the proper cross sections would then lower the impact ionization rate and slow down the progress of impact ionization, when compared to the current prediction with our code. This would, in turn, slow down diamond damage, pushing our predictions towards the experimentally observed {\it longer} disordering timescales.

We performed a test case study, in which we artificially reduced the electron inelastic scattering cross sections by a factor of two. This change slightly prolonged the electron cascading times, and delayed the onset of the damage by a few femtoseconds (see Fig.~\ref{fig:Test_peaks}, the curve marked as $\sigma/2$). However, it did not significantly reduce the temporal discrepancy between our predictions and the experiment. 

\begin{figure}[h] 
\centering 
\includegraphics[width=0.5\linewidth,trim={0 
0 15 15},clip]{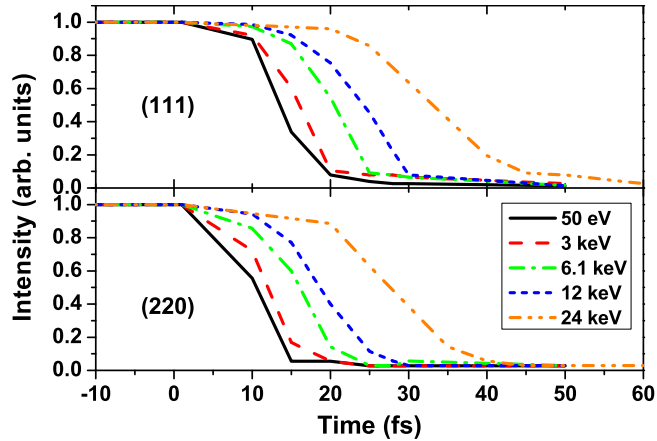}
\caption{Integrated diffraction peak intensities (111) and (220) in diamond irradiated with an X-ray pulse of 6.1 keV photon energy, 5 fs FWHM duration, at the absorbed dose of 24.9 eV/atom ($3.1 \times 10^4$ J/cm$^2$) for different photon energies.} 
\label{fig:Peaks_hw} 
\end{figure}

At this point, we can also evaluate the impact of the incoming X-ray energy on the graphitization progress. Fig.~\ref{fig:Peaks_hw} shows the decrease of intensity peaks as a function of the X-ray photon energy. Reducing the X-ray photon energy reduces the formation time of secondary electron cascades, as the corresponding energy of the photoelectron triggering the cascade is significantly lower. The transient graphitization would then still occur, however, it would start much earlier than in the reference case (X-ray energy $\sim 6.1$ keV). In the opposite case of much higher photon energies, the graphitization onset would be delayed. These observations could be used when planning any future measurements to unveil the details of this transient process.

\begin{figure}[h] 
\centering 
\includegraphics[width=0.9\linewidth,trim={25 
10 15 5},clip]{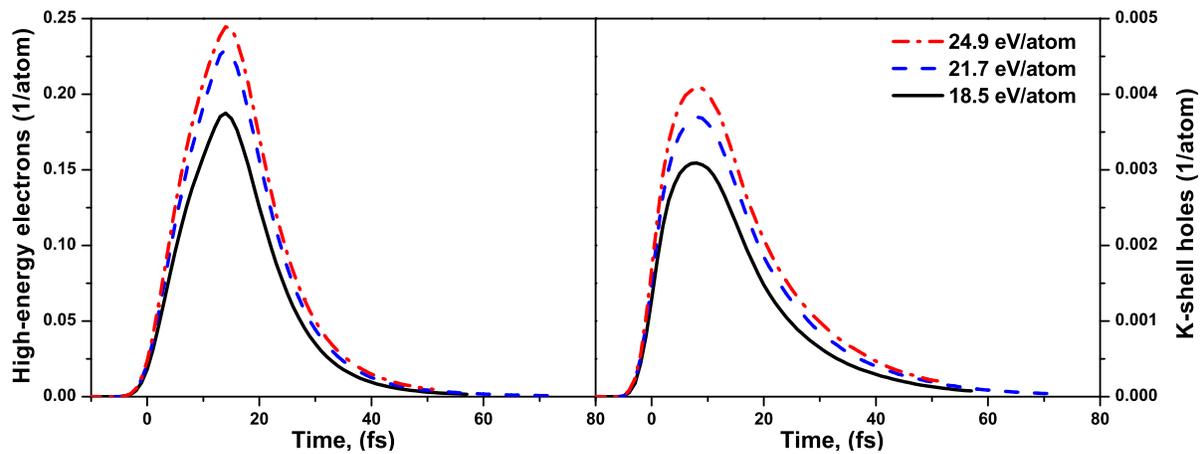} 
\caption{Fraction of high-energy electrons (left panel) and K-shell holes (right panel) in diamond created in diamond after its irradiation with X-ray pulse of 6.1 keV photon energy, 5 fs FWHM duration, at the absorbed doses of 18.5 eV/atom, 21.7 eV/atom, 24.9 eV/atom. They correspond to the experimental fluences of $2.3 \times 10^4$ J/cm$^2$, $2.7 \times 10^4$ J/cm$^2$, and $3.1 \times 10^4$ J/cm$^2$ respectively.} 
\label{fig:HEe_n_Kholes} 
\end{figure}

Secondly, the majority of the emitted high-energy electrons leave positively charged ions with K-shell holes. The presence of K-shell holes may affect the sample dynamics. Two effects can contribute: modification of the electronic band structure and the additional potential from an effective charge non-neutrality. To estimate their potential impact, we calculated the densities of K-shell holes, Fig.~\ref{fig:HEe_n_Kholes}. The calculation shows that the number of K-shell holes per atom is only on the order of $0.3-0.4 \%$ at most for the considered experimental fluences, in agreement with the estimates from Ref.~[\citeonline{Inoue2016}]. Thus, one can assume that K-shell holes do not influence the atomic dynamics significantly - in this particular case.

The number of high-energy electrons (with energies above 10 eV counted from the bottom of the conduction band) is higher than of the core holes, Fig.~\ref{fig:HEe_n_Kholes}, $\sim 18-25 \%$ per atom. Thus, we analyzed their influence on the atomic motion, performing an additional simulation with the incoming soft X-ray photons of energies $\sim 50$ eV, at the pulse parameters selected to yield exactly the same average absorbed dose per atom as the doses from the SACLA experiment. Thereby, we enforced the photoexcited and impact-ionized electrons to stay within the low energy regime only, and relax quickly to the bottom of the conduction band. In such a simulation, charge neutrality restores fast. The results of this simulation show that the electrons at the bottom of the conduction band are speeding up the atomic dynamics, rather than slowing it down - the diamond disordering in this case occurs already at $\sim 15$ fs (Fig.~\ref{fig:Peaks_hw}). Thus, we can confirm that the numerous presence of high-energy electrons seems not to be the reason for the faster atomic disordering predicted by our model, when compared to the experimental results. 

Thirdly, the XTANT model relies on the transferable tight binding parameterization, whose parameters were fitted to the equilibrium configurations of different carbon phases \cite{Xu1992}. This approximation misses the effect of the shifts of the electronic energy levels due to the presence of excited electrons. This problem is also known in the plasma community in the context of the ionization potential depression (IPD) \cite{Vinko2012a, DFT, Son2014, Ziaja2013a}. With the increasing temperature within the heated solid, higher charges appear within the sample (cf. Fig.~5 in Ref.~[\citeonline{Son2014}]). The energy levels within the band correspondingly move down. This, again, supports the expected lowering of the impact ionization rate as the ionized nuclei attract the valence electrons stronger, and the corresponding binding energies increase. Electrons occupying the valence levels below the Fermi level form attractive bonds, whereas electrons populating the levels in the conduction band above the Fermi level contribute to repulsive bonds. Thus, lowering of the conduction band levels beyond the Fermi level in the strongly heated diamond may temporally change the bonding from repulsive to attractive. This effect may 'stabilize' diamond on the way to the warm dense matter state and prolong the timescales of WDM formation. Since the  density of transiently excited electrons in diamond is close to the valence electron density in metals (see Fig.~\ref{fig:Conduction_el} below), the effects similar to the bond hardening observed in metals may be expected in diamond underway to warm dense matter state~\cite{Ernstorfer2009}.

In order to estimate the regime of applicability of the tight binding method, let us recall the observation made about the weak influence of high-energy electrons on the dynamics within the irradiated sample. As long as the fast photoexcited electrons stay within the high energy regime, i.e., are weakly coupled, they  perturb weakly the electronic structure of the solids. The tight binding method for equilibrium can still be then applied with a good accuracy, if the number of high-energy electrons is not too high to significantly affect the charge neutrality. This time span corresponds approximately to the cascading time of high-energy electrons, here $\sim 15-20$ fs (Fig.~\ref{fig:Conduction_el}). As soon as the electrons start to fill in the low-energy domain by secondary impact ionizations, the equilibrium tight-binding approximation breaks down due to the presence of so many excited electrons in the low-energy domain. 

Indepedently of these considerations, it would be good to check the influence of the specific TB parametrization \cite{Xu1992} itself on our predictions, replacing it with another one. However, a construction of another fully transferable TB parametrization or an alternative ab-initio model is a nontrivial task, which is far beyond the scope of the present work. It should be performed as a separate project. Therefore, we have to leave this point as an open question for future studies.

Finally, in these simulations, the effect of the probe pulse on the sample was neglected. Although the fluence of the probe pulse is twice as high as the fluence of the pump pulse, the probe pulse takes only 5 fs FWHM. This timescale of the diffraction measurement is too short to record any atomic relocations induced by the probe pulse and reflected by Bragg peaks \cite{ziaja2015}. We can, therefore, neglect its influence on the simulated results. 

%<-------------------------------------------------------- 
In conclusion, with our simulations, we have followed the multistep pathway of a diamond crystal exposed to intense hard X-ray FEL pulses (6.1 keV photon energy, 5 fs FWHM duration) to the warm dense matter state. Our simulations indicate a few evolution stages of the irradiated sample. Diamond transition starts with the strong electronic excitation, leading to nonthermal band gap collapse and the subsequent transient graphitization, followed by the sample disordering - within $\sim 20$ fs in total. The results are compared with the recent experimental data on diffraction peaks of diamond recorded at SACLA with the X-ray X-ray pump-probe. In particular, our simulations confirm that the intensity of (220) reflection decreases faster than the intensity of (111) reflection, in a qualitative agreement with experiment. This effect is caused by the transient graphitization of a femtosecond duration, occurring underway to warm dense matter state. Our finding then allows to explain the puzzling experimental observation on the disappearance of the diffraction peaks at different timescales.

Quantitative discrepancy between the theoretically estimated timescales of diamond disordering and the experimental ones can be attributed to the shortcomings of the tight-binding parameterization applied here under the conditions of highly excited electronic system. We estimated that the validity of this approximation under the considered experimental conditions (at average absorbed doses of 19-25 eV/atom) maintains on the timescale of up to 20 fs at most. In order to extend the applicability of our code to longer times, the TB parametrization should be replaced by another ab-initio module which can self-consistently account for changes of the electronic band structure and the impact ionization rate within such a highly excited system, out of equilibrium. Such approaches are not yet available, however, promising developments in this direction from TDDFT calculations \cite{Baczewski2016} and Hartree-Fock schemes \cite{Hao2015} are underway.

%-------------------------------------------------------- 
\section*{Methods}

\subsection{Simulation tool}
\noindent

Time evolution of x-ray irradiated diamond was modeled with the hybrid code XTANT \cite{Medvedev2013e}. The model employs tight-binding (TB) molecular dynamics (MD) to follow trajectories of atoms within a supercell. TB module also calculates transient electronic structure within the valence band and bottom of the conduction band via a direct diagonalization of the transient Hamiltonian. 
The so-called hopping integrals, entering TB Hamiltonian, are functions of interatomic distances and positions, adjusted to reproduce atomic structures of various carbonaceous materials\cite{Xu1992}. Thus, the method is capable of reproducing different phase transitions. 

This enables us to calculate the potential energy surface, $\Phi(\{ r_{ij}(t)\}, t)$, depending on all the atomic positions pairwise: 
\begin{equation}{} 
\Phi(\{ r_{ij}(t)\}, t) = \sum_{\rm i} f_e(E_{\rm i}, t) E_{\rm i} + E_{\rm rep}( \{ r_{ij} \} ) \ . 
\label{PotEn} 
\end{equation} 
Here, the first term is the attractive part formed by the electrons represented by the electron distribution function $f_e(E_i)$, populating the transient energy levels $E_i$; $E_{rep}$ is the core-core repulsive potential of ions. 

Derivative of the potential energy then enters equations of motion. They are solved for all atoms in the simulation box with periodic boundary conditions by applying Verlet algorithm in its velocity form. 

To trace the electron distribution function, $f_e(E_i)$, which enters Eq.(\ref{PotEn}), XTANT splits electrons into two fractions, according to their energy: 

(i) High-energy electrons (with energies above a threshold of $\sim 10$ eV \cite{Medvedev2013e}) are modeled as individual particles with the event-by-event Monte Carlo (MC) simulation. Their time propagation is sampled as a sequence of collisions, defined by the cross sections of interactions, which are derived from the complex dielectric function \cite{Medvedev2013e,Medvedev2015}. High-energy electrons perform secondary ionizations, exciting new electrons from the valence to the conduction band of diamond. Photoabsorption and decays of core K-shell holes are also traced within the MC scheme. 
When an electron loses its energy below the low-energy threshold, it joins the electronic low-energy fraction (see below), thereby heating up the low-energy electronic distribution. 
All the details of the MC scheme including cross sections of scattering were described in Ref.~[\citeonline{Medvedev2013e}]. 

(ii) Low-energy electrons (with energies below the threshold) are assumed to obey Fermi-Dirac distribution at all times. Their evolution is traced with a simplified Boltzmann equation \cite{Medvedev2015cc}. Boltzmann collision integral for the energy exchange with ions (electron-phonon coupling) reads (as in Ref.~ [\citeonline{Medvedev2017}]): 
\begin{eqnarray}{} 
\label{Eq:Boltzmann} 
I^{e-at}_{i,j} = w_{i,j} 
\begin{cases} 
f_e(E_i)(2 - f_e(E_j)) - f_e(E_j)(2 - f_e(E_i))G_{at}(E_i - E_j) \ , {\rm for} \ i > j , \\ 
f_e(E_i)(2 - f_e(E_j))G_{at}(E_j - E_i) - f_e(E_j)(2 - f_e(E_i)) \ , {\rm for} \ i < j ,\ 
\end{cases} 
\label{Fin_coll_int} 
\end{eqnarray} 
where $g_{at}(E)$ is the integrated Maxwellian distribution for atoms with a transient ion temperature, and $w_{i,f}$ is the electron-ion scattering rate during the actual time-step, calculated as follows \cite{Medvedev2017}: 
\begin{eqnarray}{} 
\label{Eq:Landau} 
w_{i,j} = \left| \left( \langle i(t) | j(t+\delta t) \rangle - \langle i(t +\delta t) | j(t) \rangle \right)/2 \right|^2\frac{1}{\delta t}, 
\end{eqnarray} 
where $\left<i (t)| \ \text{and} \ | j(t + \delta t) \right>$ are the $i$-th and $j$-th eigenfunctions of the Hamiltonian at the time instants $t$ and $t+\delta t$, respectively; $\delta t$ is the simulation time-step\cite{Medvedev2017}.

%-------------------------------
\subsection{Evolution of electronic subsystem in X-ray irradiated diamond}
\noindent

\begin{figure}[h] 
\centering 
\includegraphics[width=0.5\linewidth,trim={20 20 15 30},clip]{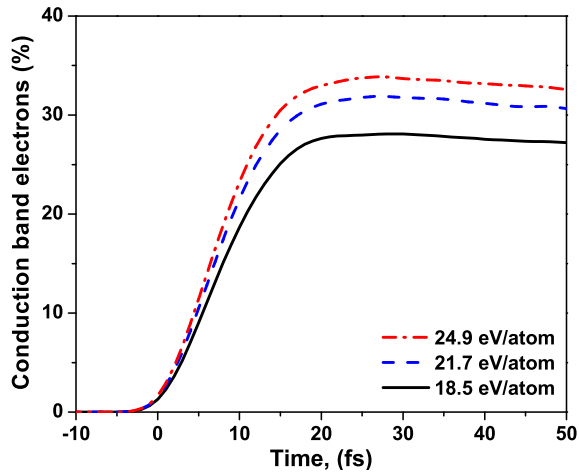} 
\caption{Fraction of conduction band electrons in diamond irradiated with an X-ray pulse of 6.1 keV photon energy, 5 fs FWHM duration, at the average absorbed doses of 18.5 eV/atom, 21.7 eV/atom, 24.9 eV/atom. They correspond to the experimental fluences of $2.3 \times 10^4$ J/cm$^2$, $2.7 \times 10^4$ J/cm$^2$, and $3.1 \times 10^4$ J/cm$^2$ respectively.} 
\label{fig:Conduction_el} 
\end{figure}

\begin{figure}[!h] 
\centering 
\includegraphics[width=0.5\linewidth,trim={20 17 15 35},clip]{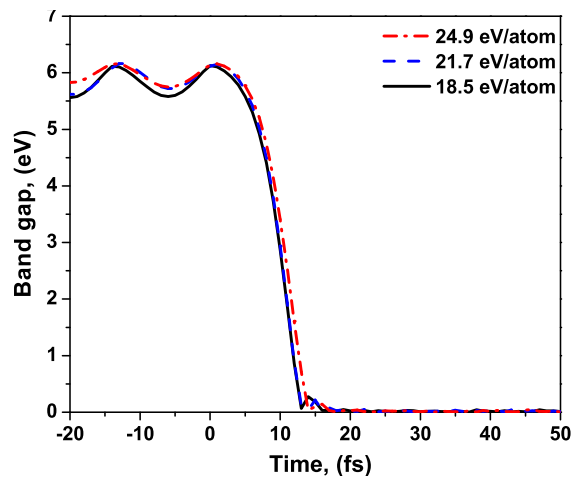} 
\caption{Band gap of diamond irradiated with an X-ray pulse of 6.1 keV photon energy, 5 fs FWHM duration, at the average absorbed doses of 18.5 eV/atom, 21.7 eV/atom, 24.9 eV/atom. They correspond to the experimental fluences of $2.3 \times 10^4$ J/cm$^2$, $2.7 \times 10^4$ J/cm$^2$, and $3.1 \times 10^4$ J/cm$^2$ respectively. } 
\label{fig:Band_gap} 
\end{figure}

Simulations of diamond damage show that this process proceeds in multiple steps, as discussed in the Introduction. Firstly, valence electrons are excited into the conduction band during and for some time after the pulse, see Fig.~\ref{fig:Conduction_el}. The number of the excited electrons is high. It overcomes the damage threshold value ($\sim 1.5$ \%)\cite{Medvedev2013e} already early during the exposure to the pump pulse \cite{Gaudin2013}. This leads to the band gap collapse at the time $\sim 15$ fs since the maximum of the pump pulse, as shown in Fig.~\ref{fig:Band_gap}. It is known from our previous studies~ \cite{Gaudin2013} that the band gap collapse indicates an onset of the irreversible phase transition, graphitization. In case of all three absorption doses considered in the experiment, the electronic processes have a similar temporal characteristics. As expected, the fraction of electrons excited increases with the increasing absorption dose.

%-------------------------------------------------------- 
\bibliography{My_Collection_2} 

\begin{thebibliography}{10}
\expandafter\ifx\csname url\endcsname\relax
  \def\url#1{\texttt{#1}}\fi
\expandafter\ifx\csname urlprefix\endcsname\relax\def\urlprefix{URL }\fi
\providecommand{\bibinfo}[2]{#2}
\providecommand{\eprint}[2][]{\url{#2}}

\bibitem{Murillo1998}
\bibinfo{author}{Murillo, M.~S.} \& \bibinfo{author}{Weisheit, J.~C.}
\newblock \bibinfo{title}{{Dense plasmas, screened interactions, and atomic
  ionization}}.
\newblock \emph{\bibinfo{journal}{Physical Reports}}
  \textbf{\bibinfo{volume}{302}}, \bibinfo{pages}{1} (\bibinfo{year}{1998}).

\bibitem{Zwanzig2001}
\bibinfo{author}{Zwanzig, R.}
\newblock \emph{\bibinfo{title}{{Nonequilibrium statistical mechanics}}}
  (\bibinfo{publisher}{Oxford University Press}, \bibinfo{year}{2001}).
\newblock
  \urlprefix\url{https://global.oup.com/academic/product/nonequilibrium-statistical-mechanics-9780195140187?cc=de{\&}lang=en{\&}}.

\bibitem{Birdsall2005}
\bibinfo{author}{Birdsall, C.~K.} \& \bibinfo{author}{Langdon, A.~B.}
\newblock \emph{\bibinfo{title}{{Plasma physics via computer simulation}}}
  (\bibinfo{publisher}{Institute of Physics Pub}, \bibinfo{year}{2005}).

\bibitem{Parr1980}
\bibinfo{author}{Parr, R.~G.}
\newblock \bibinfo{title}{{Density Functional Theory of Atoms and Molecules}}.
\newblock In \emph{\bibinfo{booktitle}{Horizons of Quantum Chemistry}},
  \bibinfo{pages}{5--15} (\bibinfo{publisher}{Springer Netherlands},
  \bibinfo{address}{Dordrecht}, \bibinfo{year}{1980}).
\newblock
  \urlprefix\url{http://www.springerlink.com/index/10.1007/978-94-009-9027-2{\_}2}.

\bibitem{Marques2012}
\bibinfo{editor}{Marques, M.~A.}, \bibinfo{editor}{Maitra, N.~T.},
  \bibinfo{editor}{Nogueira, F.~M.}, \bibinfo{editor}{Gross, E.} \&
  \bibinfo{editor}{Rubio, A.} (eds.) \emph{\bibinfo{title}{{Fundamentals of
  Time-Dependent Density Functional Theory}}}, vol. \bibinfo{volume}{837} of
  \emph{\bibinfo{series}{Lecture Notes in Physics}}
  (\bibinfo{publisher}{Springer Berlin Heidelberg}, \bibinfo{address}{Berlin,
  Heidelberg}, \bibinfo{year}{2012}).
\newblock \urlprefix\url{http://link.springer.com/10.1007/978-3-642-23518-4}.

\bibitem{Graziani2014}
\bibinfo{author}{Graziani, F.}, \bibinfo{author}{Desjarlais, M.~P.},
  \bibinfo{author}{Redmer, R.} \& \bibinfo{author}{Trickey, S.~B.}
\newblock \emph{\bibinfo{title}{{Frontiers and Challenges in Warm Dense
  Matter}}} (\bibinfo{publisher}{Springer-Verlag New York Inc},
  \bibinfo{address}{New York}, \bibinfo{year}{2014}).

\bibitem{Redmer2008}
\bibinfo{author}{Redmer, R.} \emph{et~al.}
\newblock \bibinfo{title}{{Giant planets as laboratory for high energy density
  physics}}.
\newblock In \emph{\bibinfo{booktitle}{2008 IEEE 35th International Conference
  on Plasma Science}}, \bibinfo{pages}{1--1} (\bibinfo{publisher}{IEEE},
  \bibinfo{year}{2008}).
\newblock \urlprefix\url{http://ieeexplore.ieee.org/document/4590589/}.

\bibitem{Valenza2016}
\bibinfo{author}{Valenza, R.~A.} \& \bibinfo{author}{Seidler, G.~T.}
\newblock \bibinfo{title}{{Warm dense crystallography}}.
\newblock \emph{\bibinfo{journal}{Physical Review B - Condensed Matter and
  Materials Physics}} \textbf{\bibinfo{volume}{93}}, \bibinfo{pages}{1--7}
  (\bibinfo{year}{2016}).

\bibitem{Fletcher2015}
\bibinfo{author}{Fletcher, L.~B.} \emph{et~al.}
\newblock \bibinfo{title}{{Ultrabright X-ray laser scattering for dynamic warm
  dense matter physics}}.
\newblock \emph{\bibinfo{journal}{Nature Photonics}}
  \textbf{\bibinfo{volume}{9}}, \bibinfo{pages}{274--279}
  (\bibinfo{year}{2015}).
\newblock \urlprefix\url{http://dx.doi.org/10.1038/nphoton.2015.41}.

\bibitem{Ackermann2007}
\bibinfo{author}{Ackermann, W.} \emph{et~al.}
\newblock \bibinfo{title}{{Operation of a free-electron laser from the extreme
  ultraviolet to the water window}}.
\newblock \emph{\bibinfo{journal}{Nature Photonics}}
  \textbf{\bibinfo{volume}{1}}, \bibinfo{pages}{336--342}
  (\bibinfo{year}{2007}).
\newblock \urlprefix\url{http://dx.doi.org/10.1038/nphoton.2007.76}.

\bibitem{Bostedt2016}
\bibinfo{author}{Bostedt, C.} \emph{et~al.}
\newblock \bibinfo{title}{{Linac Coherent Light Source: The first five years}}.
\newblock \emph{\bibinfo{journal}{Reviews of Modern Physics}}
  \textbf{\bibinfo{volume}{88}}, \bibinfo{pages}{015007}
  (\bibinfo{year}{2016}).
\newblock
  \urlprefix\url{http://journals.aps.org/rmp/abstract/10.1103/RevModPhys.88.015007}.

\bibitem{Pile2011}
\bibinfo{author}{Pile, D.}
\newblock \bibinfo{title}{{X-rays: First light from SACLA}}.
\newblock \emph{\bibinfo{journal}{Nature Photonics}}
  \textbf{\bibinfo{volume}{5}}, \bibinfo{pages}{456--457}
  (\bibinfo{year}{2011}).
\newblock
  \urlprefix\url{http://www.nature.com/doifinder/10.1038/nphoton.2011.178}.

\bibitem{Ganter2011}
\bibinfo{editor}{Ganter, R.} \emph{et~al.} (eds.)
  \emph{\bibinfo{title}{{SwissFEL Conceptual Design Report}}}
  (\bibinfo{publisher}{PSI Bericht}, \bibinfo{address}{Villigen},
  \bibinfo{year}{2011}), \bibinfo{edition}{v19} edn.
\newblock \urlprefix\url{www.psi.ch/swissfel}.

\bibitem{Altarelli2011}
\bibinfo{author}{Altarelli, M.}
\newblock \bibinfo{title}{{The European X-ray free-electron laser facility in
  Hamburg}}.
\newblock \emph{\bibinfo{journal}{Nuclear Instruments and Methods in Physics
  Research Section B: Beam Interactions with Materials and Atoms}}
  \textbf{\bibinfo{volume}{269}}, \bibinfo{pages}{2845--2849}
  (\bibinfo{year}{2011}).
\newblock
  \urlprefix\url{http://linkinghub.elsevier.com/retrieve/pii/S0168583X11003855}.

\bibitem{Lorazo2006}
\bibinfo{author}{Lorazo, P.}, \bibinfo{author}{Lewis, L.} \&
  \bibinfo{author}{Meunier, M.}
\newblock \bibinfo{title}{{Thermodynamic pathways to melting, ablation, and
  solidification in absorbing solids under pulsed laser irradiation}}.
\newblock \emph{\bibinfo{journal}{Physical Review B}}
  \textbf{\bibinfo{volume}{73}}, \bibinfo{pages}{134108}
  (\bibinfo{year}{2006}).
\newblock \urlprefix\url{http://link.aps.org/doi/10.1103/PhysRevB.73.134108}.

\bibitem{Zastrau2012}
\bibinfo{author}{Zastrau, U.} \emph{et~al.}
\newblock \bibinfo{title}{{XUV spectroscopic characterization of warm dense
  aluminum plasmas generated by the free-electron-laser FLASH}}.
\newblock \emph{\bibinfo{journal}{Laser and Particle Beams}}
  \textbf{\bibinfo{volume}{30}}, \bibinfo{pages}{45--56}
  (\bibinfo{year}{2012}).
\newblock
  \urlprefix\url{http://www.journals.cambridge.org/abstract{\_}S026303461100067X}.

\bibitem{Medvedev2015}
\bibinfo{author}{Medvedev, N.}, \bibinfo{author}{Tkachenko, V.} \&
  \bibinfo{author}{Ziaja, B.}
\newblock \bibinfo{title}{{Modeling of Nonthermal Solid-to-Solid Phase
  Transition in Diamond Irradiated with Femtosecond x-ray FEL Pulse}}.
\newblock \emph{\bibinfo{journal}{Contributions to Plasma Physics}}
  \textbf{\bibinfo{volume}{55}}, \bibinfo{pages}{12--34}
  (\bibinfo{year}{2015}).
\newblock \urlprefix\url{http://doi.wiley.com/10.1002/ctpp.201400026}.

\bibitem{Ziaja2008}
\bibinfo{author}{Ziaja, B.}, \bibinfo{author}{Wabnitz, H.},
  \bibinfo{author}{Weckert, E.} \& \bibinfo{author}{M{\"{o}}ller, T.}
\newblock \bibinfo{title}{{Femtosecond non-equilibrium dynamics of clusters
  irradiated with short intense VUV pulses}}.
\newblock \emph{\bibinfo{journal}{New Journal of Physics}}
  \textbf{\bibinfo{volume}{10}}, \bibinfo{pages}{043003}
  (\bibinfo{year}{2008}).
\newblock \urlprefix\url{http://stacks.iop.org/1367-2630/10/i=4/a=043003}.

\bibitem{Inoue2016}
\bibinfo{author}{Inoue, I.} \emph{et~al.}
\newblock \bibinfo{title}{{Observation of femtosecond X-ray interactions with
  matter using an X-ray-X-ray pump-probe scheme.}}
\newblock \emph{\bibinfo{journal}{Proceedings of the National Academy of
  Sciences of the United States of America}} \textbf{\bibinfo{volume}{113}},
  \bibinfo{pages}{1492--7} (\bibinfo{year}{2016}).
\newblock \urlprefix\url{http://www.ncbi.nlm.nih.gov/pubmed/26811449}.

\bibitem{Henke1993}
\bibinfo{author}{Henke, B.}, \bibinfo{author}{Gullikson, E.} \&
  \bibinfo{author}{Davis, J.}
\newblock \bibinfo{title}{{X-Ray Interactions: Photoabsorption, Scattering,
  Transmission, and Reflection at E = 50-30,000 eV, Z = 1-92}}.
\newblock \emph{\bibinfo{journal}{Atomic Data and Nuclear Data Tables}}
  \textbf{\bibinfo{volume}{54}}, \bibinfo{pages}{181--342}
  (\bibinfo{year}{1993}).
\newblock \urlprefix\url{http://dx.doi.org/10.1006/adnd.1993.1013}.

\bibitem{Dharma-wardana2016a}
\bibinfo{author}{Dharma-wardana, M.}
\newblock \bibinfo{title}{{Current Issues in Finite-T Density-Functional Theory
  and Warm-Correlated Matter †}}.
\newblock \emph{\bibinfo{journal}{Computation}} \textbf{\bibinfo{volume}{4}},
  \bibinfo{pages}{16} (\bibinfo{year}{2016}).
\newblock \urlprefix\url{http://www.mdpi.com/2079-3197/4/2/16{\%}0A}.

\bibitem{Siders1999}
\bibinfo{author}{Siders, C.~W.}
\newblock \bibinfo{title}{{Detection of Nonthermal Melting by Ultrafast X-ray
  Diffraction}}.
\newblock \emph{\bibinfo{journal}{Science}} \textbf{\bibinfo{volume}{286}},
  \bibinfo{pages}{1340--1342} (\bibinfo{year}{1999}).
\newblock
  \urlprefix\url{http://www.sciencemag.org/content/286/5443/1340.abstract}.

\bibitem{Sundaram2002}
\bibinfo{author}{Sundaram, S.~K.} \& \bibinfo{author}{Mazur, E.}
\newblock \bibinfo{title}{{Inducing and probing non-thermal transitions in
  semiconductors using femtosecond laser pulses.}}
\newblock \emph{\bibinfo{journal}{Nature materials}}
  \textbf{\bibinfo{volume}{1}}, \bibinfo{pages}{217--24}
  (\bibinfo{year}{2002}).
\newblock \urlprefix\url{http://dx.doi.org/10.1038/nmat767}.

\bibitem{Medvedev2013e}
\bibinfo{author}{Medvedev, N.}, \bibinfo{author}{Jeschke, H.~O.} \&
  \bibinfo{author}{Ziaja, B.}
\newblock \bibinfo{title}{{Nonthermal phase transitions in semiconductors
  induced by a femtosecond extreme ultraviolet laser pulse}}.
\newblock \emph{\bibinfo{journal}{New Journal of Physics}}
  \textbf{\bibinfo{volume}{15}}, \bibinfo{pages}{015016}
  (\bibinfo{year}{2013}).
\newblock \urlprefix\url{http://stacks.iop.org/1367-2630/15/i=1/a=015016}.

\bibitem{Medvedev2017}
\bibinfo{author}{Medvedev, N.}, \bibinfo{author}{Li, Z.},
  \bibinfo{author}{Tkachenko, V.} \& \bibinfo{author}{Ziaja, B.}
\newblock \bibinfo{title}{{Electron-ion coupling in semiconductors beyond
  Fermi's golden rule}}.
\newblock \emph{\bibinfo{journal}{Physical Review B}}
  \textbf{\bibinfo{volume}{95}}, \bibinfo{pages}{014309}
  (\bibinfo{year}{2017}).
\newblock \urlprefix\url{http://link.aps.org/doi/10.1103/PhysRevB.95.014309}.

\bibitem{Medvedev2013f}
\bibinfo{author}{Medvedev, N.}, \bibinfo{author}{Jeschke, H.~O.} \&
  \bibinfo{author}{Ziaja, B.}
\newblock \bibinfo{title}{{Nonthermal graphitization of diamond induced by a
  femtosecond x-ray laser pulse}}.
\newblock \emph{\bibinfo{journal}{Physical Review B}}
  \textbf{\bibinfo{volume}{88}}, \bibinfo{pages}{224304}
  (\bibinfo{year}{2013}).
\newblock \urlprefix\url{http://link.aps.org/doi/10.1103/PhysRevB.88.224304}.

\bibitem{Gaudin2013}
\bibinfo{author}{Gaudin, J.} \emph{et~al.}
\newblock \bibinfo{title}{{Photon energy dependence of graphitization threshold
  for diamond irradiated with an intense XUV FEL pulse}}.
\newblock \emph{\bibinfo{journal}{Physical Review B}}
  \textbf{\bibinfo{volume}{88}}, \bibinfo{pages}{060101(R)}
  (\bibinfo{year}{2013}).
\newblock \urlprefix\url{http://link.aps.org/doi/10.1103/PhysRevB.88.060101}.

\bibitem{Tavella2017}
\bibinfo{author}{Tavella, F.} \emph{et~al.}
\newblock \bibinfo{title}{{Soft x-ray induced femtosecond solid-to-solid phase
  transition}}.
\newblock \emph{\bibinfo{journal}{High Energy Density Physics}}
  \textbf{\bibinfo{volume}{24}}, \bibinfo{pages}{22--27}
  (\bibinfo{year}{2017}).
\newblock
  \urlprefix\url{http://www.sciencedirect.com/science/article/pii/S1574181817300617}.

\bibitem{reciprOgraph}
\bibinfo{author}{Hardaker, W.}, \bibinfo{author}{Schoeni, N.},
  \bibinfo{author}{Chapuis, G.}, \bibinfo{author}{Casademont, N.} \&
  \bibinfo{author}{Sisto, M.}
\newblock \bibinfo{title}{{ Available online at
  http://escher.epfl.ch/reciprOgraph/}}.
\newblock \urlprefix\url{{ http://escher.epfl.ch/reciprOgraph/ }}.

\bibitem{Jeschke2001}
\bibinfo{author}{Jeschke, H.}, \bibinfo{author}{Garcia, M.} \&
  \bibinfo{author}{Bennemann, K.}
\newblock \bibinfo{title}{{Theory for the Ultrafast Ablation of Graphite
  Films}}.
\newblock \emph{\bibinfo{journal}{Physical Review Letters}}
  \textbf{\bibinfo{volume}{87}}, \bibinfo{pages}{015003}
  (\bibinfo{year}{2001}).
\newblock
  \urlprefix\url{http://link.aps.org/doi/10.1103/PhysRevLett.87.015003}.

\bibitem{Medvedev2015cc}
\bibinfo{author}{Medvedev, N.}
\newblock \bibinfo{title}{{X-ray-induced electron cascades in dielectrics
  modeled with XCASCADE code: Effect of impact ionization cross sections}}.
\newblock In \emph{\bibinfo{booktitle}{Proceedings of SPIE - The International
  Society for Optical Engineering}}, vol. \bibinfo{volume}{9511}
  (\bibinfo{year}{2015}).

\bibitem{Son2014}
\bibinfo{author}{Son, S.-K.}, \bibinfo{author}{Thiele, R.},
  \bibinfo{author}{Jurek, Z.}, \bibinfo{author}{Ziaja, B.} \&
  \bibinfo{author}{Santra, R.}
\newblock \bibinfo{title}{Quantum mechanical calculation of
  ionization-potential lowering in dense plasmas}.
\newblock \emph{\bibinfo{journal}{Phys. Rev. X}} \textbf{\bibinfo{volume}{4}},
  \bibinfo{pages}{031004} (\bibinfo{year}{2014}).

\bibitem{Xu1992}
\bibinfo{author}{Xu, C.~H.}, \bibinfo{author}{Wang, C.~Z.},
  \bibinfo{author}{Chan, C.~T.} \& \bibinfo{author}{Ho, K.~M.}
\newblock \bibinfo{title}{{A transferable tight-binding potential for carbon}}.
\newblock \emph{\bibinfo{journal}{Journal of Physics: Condensed Matter}}
  \textbf{\bibinfo{volume}{4}}, \bibinfo{pages}{6047--6054}
  (\bibinfo{year}{1992}).
\newblock \urlprefix\url{http://stacks.iop.org/0953-8984/4/i=28/a=006}.

\bibitem{Vinko2012a}
\bibinfo{author}{Vinko, S.~M.} \emph{et~al.}
\newblock \bibinfo{title}{{Creation and diagnosis of a solid-density plasma
  with an X-ray free-electron laser.}}
\newblock \emph{\bibinfo{journal}{Nature}} \textbf{\bibinfo{volume}{482}},
  \bibinfo{pages}{59--62} (\bibinfo{year}{2012}).
\newblock \urlprefix\url{http://dx.doi.org/10.1038/nature10746
  http://www.ncbi.nlm.nih.gov/pubmed/22278059}.

\bibitem{DFT}
\bibinfo{author}{Vinko, S.}, \bibinfo{author}{Ciricosta, O.} \&
  \bibinfo{author}{Wark, J.}
\newblock \bibinfo{title}{{Density functional theory calculations of continuum
  lowering in strongly coupled plasmas}}.
\newblock \emph{\bibinfo{journal}{Nature Communications}}
  \textbf{\bibinfo{volume}{5}}, \bibinfo{pages}{3533} (\bibinfo{year}{2014}).

\bibitem{Ziaja2013a}
\bibinfo{author}{Ziaja, B.} \emph{et~al.}
\newblock \bibinfo{title}{{Photoelectron spectroscopy method to reveal
  ionization potential lowering in nanoplasmas}}.
\newblock \emph{\bibinfo{journal}{Journal of Physics B: Atomic, Molecular and
  Optical Physics}} \textbf{\bibinfo{volume}{46}}, \bibinfo{pages}{164009}
  (\bibinfo{year}{2013}).
\newblock \urlprefix\url{http://stacks.iop.org/0953-4075/46/i=16/a=164009}.

\bibitem{Ernstorfer2009}
\bibinfo{author}{Ernstorfer, R.} \emph{et~al.}
\newblock \bibinfo{title}{{The formation of warm dense matter: experimental
  evidence for electronic bond hardening in gold.}}
\newblock \emph{\bibinfo{journal}{Science (New York, N.Y.)}}
  \textbf{\bibinfo{volume}{323}}, \bibinfo{pages}{1033--7}
  (\bibinfo{year}{2009}).
\newblock \urlprefix\url{http://www.ncbi.nlm.nih.gov/pubmed/19164708}.

\bibitem{ziaja2015}
\bibinfo{author}{Ziaja, B.}, \bibinfo{author}{Medvedev, N.},
  \bibinfo{author}{Tkachenko, V.}, \bibinfo{author}{Maltezopoulos, T.} \&
  \bibinfo{author}{Wurth, W.}
\newblock \bibinfo{title}{{Time-resolved observation of band-gap shrinking and
  electron-lattice thermalization within X-ray excited gallium arsenide}}.
\newblock \emph{\bibinfo{journal}{Scientific Reports}}
  \textbf{\bibinfo{volume}{5}}, \bibinfo{pages}{18068} (\bibinfo{year}{2015}).
\newblock
  \urlprefix\url{http://www.nature.com/srep/2015/151211/srep18068/full/srep18068.html}.
\newblock \eprint{/arxiv.org/abs/1510.00610}.

\bibitem{Baczewski2016}
\bibinfo{author}{Baczewski, A.}, \bibinfo{author}{Shulenburger, L.},
  \bibinfo{author}{Desjarlais, M.}, \bibinfo{author}{Hansen, S.} \&
  \bibinfo{author}{Magyar, R.}
\newblock \bibinfo{title}{{X-ray Thomson Scattering in Warm Dense Matter
  without the Chihara Decomposition}}.
\newblock \emph{\bibinfo{journal}{Physical Review Letters}}
  \textbf{\bibinfo{volume}{116}}, \bibinfo{pages}{115004}
  (\bibinfo{year}{2016}).
\newblock
  \urlprefix\url{http://link.aps.org/doi/10.1103/PhysRevLett.116.115004}.

\bibitem{Hao2015}
\bibinfo{author}{Hao, Y.}, \bibinfo{author}{Inhester, L.},
  \bibinfo{author}{Hanasaki, K.}, \bibinfo{author}{Son, S.-K.} \&
  \bibinfo{author}{Santra, R.}
\newblock \bibinfo{title}{{Efficient electronic structure calculation for
  molecular ionization dynamics at high x-ray intensity}}.
\newblock \emph{\bibinfo{journal}{Structural Dynamics}}
  \textbf{\bibinfo{volume}{2}}, \bibinfo{pages}{041707} (\bibinfo{year}{2015}).
\newblock \urlprefix\url{http://aca.scitation.org/doi/10.1063/1.4919794}.

\end{thebibliography}

%-------------------------------------------------------- 
\section*{Acknowledgements} 

NM and BZ thank Ichiro Inoue, Robin Santra, Ilme Schlichting, Sang-Kil Son, Victor Tkachenko, Makina Yabashi, and Hitoki Yoneda for helpful discussions. In particular, BZ thanks Ilme Schlichting for the inspiring communication which stimulated this study. Partial financial support from the Czech Ministry of Education (Grants LG15013 and LM2015083) is acknowledged by N. Medvedev. 

%-------------------------------------------------------- 
\section*{Author contributions statement} 

N.M. performed the calculations, N.M. and B.Z. analyzed and interpreted the results. All authors contributed to writing the manuscript. 

%-------------------------------------------------------- 
\section*{Additional information} 

The authors declare no competing interests. 

%-------------------------------------------------------- 

\end{document}